\documentclass[12pt,a4paper]{article}
\pdfoutput=1
\usepackage{amsmath,amsfonts,amscd,amssymb,cite}
\usepackage[utf8]{inputenc}
\usepackage[english]{babel}
\usepackage[T2A]{fontenc}
\usepackage{upref}
\usepackage{xspace}
\usepackage{enumerate}
\usepackage{amsthm}
\usepackage{amsxtra}
\usepackage{array}
\usepackage[pdftex]{graphicx}
\usepackage{xcolor}

\definecolor{linkcolor}{HTML}{FF0000}
\definecolor{urlcolor}{HTML}{0000CC}
\usepackage{hyperref}
\hypersetup{pdfstartview=FitH,  linkcolor=linkcolor, urlcolor=urlcolor, colorlinks=true}

\let\cal\mathcal

\DeclareSymbolFont{bletters}{OML}{cmm}{bx}{it}
\DeclareMathSymbol{\bla}{\mathord}{bletters}{'025}
\DeclareMathSymbol{\bmu}{\mathord}{bletters}{'026}
\DeclareMathSymbol{\bth}{\mathord}{bletters}{'022}
\DeclareMathSymbol{\bfI}{\mathord}{bletters}{"49}
\DeclareMathSymbol{\bdl}{\mathord}{bletters}{"0E}
\DeclareMathSymbol{\bDl}{\mathord}{bletters}{"001}

\def \bdl{\boldsymbol\dl}

\def \bsi{\boldsymbol\si}
\def \bta{\boldsymbol\eta}
\def \bxi{\boldsymbol\xi}

\def \bvphi{\boldsymbol\varphi}

\def \bx{\bf{x}}

\def \bsi{\boldsymbol\si}

\def \be{\beta}
\def \ga{\gamma}
\def \dl{\delta}

\def \si{\sigma}
\def \om{\omega}

\def \Dl{\Delta}
\def \La{\Lambda}

\def \ph{\varphi}

\def \cS{\cal S}
\def \cZ{\cal Z}
\def \cB{\cal B}

\def \CU{\mathcal U}

\def \cC{\cal C}

\def \cE{\cal E}

\def \cK{\cal K}
\def \cN{\cal N}
\def \cP{\cal P}

\def \cS{\cal S}

\def \cZ{\cal Z}

\def \cL{{\cal L}}
\def \CU{{\cal U}}
\def \CW{{\cal W}}

\def \c{{\rm c}\,}
\def \m{{\rm m}\,}
\def \d{{\rm d}}

\def \nt{{\widetilde n}}

\def \r{\rangle}
\def \l{\langle}
\def \lav{\l\!\l}
\def \rav{\r\!\r}

\def \1{^{-1}}
\def \cd{\partial}

\begin{document}
\title {
$${}$$
{\bf The Einstein-like field theory
and the dislocations with finite-sized core}}
\author{
$${}$$
{\bf\Large Cyril Malyshev}\\
$${}$$\\
{\it\small Steklov Mathematical Institute}
{\it\small (St.-Petersburg Department)}\\
{\it\small Fontanka 27, St.-Petersburg, 191023, RUSSIA}}

\date{}

\maketitle

\begin{abstract}
\noindent
Einstein-like Lagrangian field theory is
developed to describe elastic solid
containing dislocations with
finite-sized core. The framework of the Riemann--Cartan geometry in three
dimensions is used, and the core self-energy is expressed by the translational part of the
general Lagrangian quadratic in torsion and curvature. In the Hilbert--Einstein case,
the gauge equation plays the role of non-conventional incompatibility law. The
stress tensor of the modified screw
dislocations is smoothed out within the
core. The renormalization of the elastic
constants caused by proliferation of the dislocation dipoles
is considered. The use of the
singularityless dislocation solution
modifies the renormalization of the shear
modulus in comparison with the case of
singular dislocations.
\end{abstract}

{\bf\small Keywords:} {\small Translational gauging;
Hilbert-Einstein gauge equation;
Screw dislocation; Shear modulus.}

\thispagestyle{empty}
\newpage

\section{Introduction}\label{cor:sec1}

Considerable role of geometry in modern
theoretical physics is widely acknowledged. The
differential geometry provides a
framework for analogies between the
low-di\-men\-si\-on\-al gravity and the defects in
solids \cite{kl11, kl12, mult, kat, mok, mok1, katmeth}. Geometric considerations are of importance for investigation of interplay between such subjects as the statistical physics of defects, symmetries, and phase transitions \cite{nelson, hugh, liu}. The multivalued fields are of great
importance in the condensed matter physics for
description of defects and phase transitions,
\cite{mult}. Multivaluedness of the
transformation functions (of the
displacement field in the case of dislocated
crystals) is responsible for the topological
non-triviality of the line-like defects. The
point singularities of these functions
constitute the densities of the line-like
defects. The multivalued
coordinate transformations (corresponding to
the defected crystal) transform a flat space
into general affine spaces with curvature
and torsion. It is the Riemann-Cartan
geometry of the manifolds with curvature and
torsion which is relevant to the theory of
solids with dislocations and disclinations,
\cite{kl11, kl12, kat, mok, mok1, mult, katmeth}.\footnote{Influence of, so-called, `emergent geometry'
on fermions in presence of elastic strains is discussed in \cite{vol1}.}

An original gauge approach to the statistical physics of the line-like defects (of dislocations and disclinations, in particular) is summarized in the monographs \cite{kl11, kl12, mult}.
The dislocation core energy is proposed in \cite{kl12} in order to formulate the lattice model of the defect melting and the corresponding disorder field theory. The dislocation core energy in \cite{kl12} depends quadratically on the defect tensor which is expressed
through the disclination and dislocation densities.

The singular character of the dislocation solutions of incompatible elasticity theory is an idealization since the dislocation core is not captured by the classical elasticity. The Lagrangian gauge approaches proposed in \cite{val, ed1, laz2, def3} enable to
describe the dislocations possessing the cores of finite size.\footnote{
An interest to non-singular strings, \cite{str1}, to fluxes of finite width, \cite{str2},
and to topological media characterized by flat bands in cores of defects, \cite{vol2}, also deserves a mention.} The gauge Lagrangians are chosen in \cite{val, ed1, laz2} in the form
quadratic in the gauge field strength, i.e., in the dislocation density. The approach of the Einstein-like field theory \cite{def3, def4} is based on the Hilbert--Einstein gauge Lagrangian for description of the core self-energy.
Remind that the geometric theory of defects proposed in \cite{kat} is based on the most general eight-parameter gauge Lagrangian invariant with respect of localized action of three-dimensional Euclidean group. Since the dislocation density is identified as the differential-geometric torsion, \cite{kl12, mult, katmeth}, the gauge Lagrangians \cite{val, ed1, laz2, def3, def4} may be viewed as parts of the gauge Lagrangian \cite{kat}.

In the case of the
screw dislocations, the Einstein-type gauge
equation plays the role of non-conventional
incompatibility law. The model \cite{def4} allows for a
continuation of the stresses of
the screw dislocation within the core so
that artificial cut-off does not occur both in linear and quadratic approximations. The dislocation cross-section
is no longer ``point-wise'', while sufficiently
far from the core region the dislocation demonstrates its
topological nature.

A thermodynamical description is developed in \cite{def5, aopd} for non-singular screw
dislocations described by the Einstein-like field theory \cite{def4}. The dislocation core energy in \cite{def5, aopd} originates, by means of linearizations, from the Lagrangian \cite{def3}.
The partition function of
collection
of the screw dislocations
is proposed in \cite{def5} in the functional integral form so that
the modified screw dislocation arises as its saddle point. Eventually, the effective energy of the system of positive and negative modified dislocations along an elastic cylinder is that of two-component two-dimensional Coulomb gas of particles with smoothed out coupling.\footnote{The approach \cite{def5, aopd} is concerned with individual defects, and it is not of the type of the disorder field theories developed in \cite{kl11, kl12, mult} for various phase transitions (e.g., superfluidity, superconductivity, melting).
}

Proliferation of the
dislocation dipoles is responsible for the renormalization of the elastic constants, \cite{holz, nel1, nel12, nel13}. From the viewpoint of nano-physics it is attractive, \cite{grun, rab}, to study the properties of the elastic moduli using the modified dislocation solutions
\cite{val, ed1, laz2, def3, def4, ga}. In Refs.~\cite{def5, aopd} the
renormalization of the elastic constants due
to proliferation of dipoles of the modified screw
dislocations is studied. The renormalization of
the shear modulus under the influence of the dislocation cores
is obtained near the melting transition \cite{aopd}.

\section{The dislocations with finite-sized core}
\label{cor:sec2}

Initial and deformed states of the
dislocated three-dimensional solid are
described by squared length elements $g_{i j}
{\d}x^i {\d}x^j$ and $\eta_{a b} {\d}\xi^a {\d}\xi^b$, provided that the map ${\bf x}\,\longmapsto\,\bxi ({\bf x})$ transforms initial state to the deformed one. The approach of \cite{def4} is based on the Eulerian picture expressed by the strain tensor $e_{a b}$ referred to the deformed state:
\begin{equation}
\eta_{a b} {\d}\xi^a {\d}\xi^b-g_{i j} {\d}x^i {\d}x^j= 2 e_{a b} {\d}\xi^a {\d}\xi^b
\,, \label{1}
\end{equation}
where
\begin{equation}
2 e_{a b}\equiv \eta_{a b} - g_{a b},
\qquad g_{a b}\equiv g_{i j} {\cB}_a^{\,\,i}{\cB}_b^{\,\,j}\,. \label{cor:2}
\end{equation}
The metric $g_{a b}$ (\ref{cor:2}) is the Cauchy deformation tensor, and the components ${\cB}_a^{\,\,i}$ are given by
one-forms ${\d}x^i={\cB}_a^{\,\,i} {\d}\xi^a$.
Assume that ${\cB}_a^{\,\,i}$ are ${\sf
T}(3)$-gauged:
\begin{equation}
{\cB}_a^{\,\,i}\,=\,\frac{\cd x^i}{\cd\xi^a}
\,-\,\ph_a^{\,\,i}\,.
\label{cor:5}
\end{equation}
Then, dislocations are allowed provided that one-forms ${\cB}_a^{\,\,i} {\d}\xi^a$ are not globally closed
due to the gauge potentials $\ph_a^{\,\,i}$. The entries $\ph_a^{\,\,i}$ are the translational gauge
potentials, which behave under the local shifts
$x^i\,\longrightarrow\,x^i+\eta^i(x)$ as follows:
\begin{equation}
\begin{array}{lcr}
\displaystyle{
\frac{\cd x^i}{\cd\xi^a}}&\longrightarrow&\displaystyle{
\frac{\cd x^j}{\cd\xi^a}\,\Bigl(\dl^i_j\,+\,\frac{\cd \eta^i}{\cd x^j}
\Bigr)}\,,\\
[0.5cm]
\displaystyle{
\ph_a^{\,\,i}}&\longrightarrow&\displaystyle{\!\!\!\!\!\!\!
\ph_a^{\,\,i}\,+\,\frac{\cd x^j}{\cd\xi^a}\,
\frac{\cd \eta^i}{\cd x^j}}\,.
\end{array}
\label{cor:6}
\end{equation}
The transformations (\ref{cor:6}) ensure the gauge invariance of ${\cB}_a^{\,\,i}$
(\ref{cor:5}). Non-triviality of the
core for the screw dislocation should be associated with the gauge fields ${\bvphi}\equiv (\ph_a^{\,\,i})$.

The Lagrangian of the model includes, apart from
elastic energy, the trans\-lat\-ion\-al
part of the eight-parameter Lagrangian
${\cL}_{\rm g}(\om, \cB)$, which is invariant under
the coordinate shifts and local rotations
\cite{kat}. The translational part of
${\cL}_{\rm g}$ is given by three independent
invariants quadratic in the torsion
tensor components (identified as the dislocation density components) ${\cal
T}^{\,\,\, c}_{a b} =(\partial_a
{{\cB}_b}^{i}-\partial_b {{\cB}_a}^{i})
B_{\,i}^c$ (here $B_{\,j}^c$ are reciprocals of
${{\cB}_a}^{i}$):
\begin{equation}
{\cB}^{\1}{\cL}_{{\rm g}}\bigl|_{\omega=0}\,=\,-\,\frac {1}4 {\cal T}_{a b
c}(\be_1 {\cal T}^{a b c}\,+\,\be_2 {\cal
T}^{c a b}\,+\,\be_3 {\cal T}^{e b}_{\quad
e} \eta^{a c})\,, \quad {\cB}\equiv
\det{\cB}_a^{\,\,i}. \label{cor:lagr}
\end{equation}
The model is governed by the Hilbert--Einstein
gauge equation,
\begin{equation}
G^{ef}\,=\,\displaystyle{\frac{1}{2 \ell}}\,\bigl(\si^{ef}\,-
\,(\si_{{\rm bg}})^{ef}\bigr)\,, \label{cor:732}
\end{equation}
provided that $\be_1=-\ell$, $\be_2= 2\ell$,
$\be_3= 4\ell$ (the, so-called, Hilbert--Einstein parametrization, \cite{kat}). Left-hand side of (\ref{cor:732}) is given by the Einstein tensor $G^{e f} \equiv  \frac 14\,{\cE}^{e a b} {\cE}^{f c d} {\rm R}_{a b c d}$, where ${\cE}^{a b c}$ is the Levi-Civita tensor, and ${{\rm R}_{a b c}}^d$ is the Riemann--Christoffel tensor calculated for the metric $g_{a b}$ (\ref{cor:2}). The driving source in
right-hand side of (\ref{cor:732}) is expressed by the difference $\si^{ef} - (\si_{{\rm bg}})^{ef}$, where $\si^{ef}$ and $(\si_{{\rm bg}})^{ef}$ are the stress tensors corresponding to, so-called, total and \textit{background} contributions.
The field ${\bsi}_{{\rm bg}}\equiv (\si_{{\rm bg}})^{ef}$ is determined by a prescribed distribution of the background dislocations. Besides, $\ell$ is the energy scale of the gauge field ${\bvphi}$. The equilibrium equations are:
$\stackrel{(\eta)}{\nabla}_a \si^{a b}=0$, $\stackrel{(\eta)}{\nabla}_a (\si_{{\rm bg}})^{a b}=0$, where $\stackrel{(\eta)}{\nabla}_a$ is the covariant derivative with respect to $\eta_{a b}$.

The \textit{modified} (i.e., non-singular) screw
dislocation arises in ${\sf
T}(3)$-gauge model proposed in \cite{def4}. In the first order, the stress--strain relation
is given by $\stackrel{(1)}{\si}_{\phi z}
= 2 \mu\!\!\stackrel{(1)}{e}_{\phi z}$, while stress field of the first-order, $\stackrel{(1)}{\si}_{\phi z}$, takes the form in the cylindrical coordinates:
\begin{equation}
\stackrel{(1)}{\si}_{\phi z}\,=\,-\,\mu\,\cd_\rho\!
\stackrel{(1)}{\phi}\,=\frac{b
\mu}{2\pi}\,\frac{1}\rho \bigl(1-\kappa\rho K_1(\kappa\rho)\bigr)\,,\qquad
\stackrel{(1)}{\phi}\,\equiv\frac{- b}{2\pi}\,\bigl(\log\rho + K_0(\kappa\rho)\bigr)
\,.
\label{cor:10}
\end{equation}
Here, $\stackrel{(1)}{\phi}$ is the stress potential of the modified dislocation, $\mu$ is the shear modulus, the Burgers vector component along $z$-axis is $b_z=b$, and $\kappa=(\mu/\ell)^{\frac12}$.
The asymptotical behavior of the stress
at $\kappa\rho\gg 1$ is given by the background contribution
\[(\stackrel{(1)}\si_{{\rm bg}})_{\phi z}\,=\,\frac{b
\mu}{2\pi}\,\frac{1}\rho\,.\] The dislocation core corresponds to $\rho\,
{\stackrel{<}{_\sim}}\, 1/\kappa$, since the gauge correction to $\frac{1}{\rho}$ is exponentially small outside the core. Inside the core, $\stackrel{(1)}{\si}_{\phi z}$ behaves as ${\sf A}\rho\log({\sf B}\rho)$ at $\rho\to 0$.

In the second order, the stress components $\stackrel{(2)}{\si}_{\rho \rho}$,
$\stackrel{(2)}{\si}_{\phi \phi}$, $\stackrel{(2)}{\si}_{z z}$ differ from zero, \cite{def4}. For instance, sufficiently far from the core we obtain:
\begin{equation}
\begin{array}{rcl}
\stackrel{(2)}{\si}_{\rho \rho}\Bigl|_{\kappa\rho\gg 1} &\approx&
\displaystyle{ \frac{2 \nt}{\rho^2}
\biggl(\log\frac{\rho}{\rho_1}\,-\,
\Bigl(\frac{\rho}{R}\Bigr)^2\log \frac{R}{\rho_1}
\biggr) }\,,\\ [0.3cm]
\stackrel{(2)}{\si}_{\phi \phi}\Bigl|_{\kappa\rho\gg 1} &\approx& \displaystyle{ \frac{2 \nt}{\rho^2}
\biggl(1\,-\,\log\frac{\rho}{\rho_1}\,-\,
\Bigl(\frac{\rho}{R}\Bigr)^2\log \frac{R}{\rho_1}
\biggr) }\,,
\end{array}
\label{cor:14}
\end{equation}
where $\nt\sim \mu b^2$, and $R$ is the cylinder's external radius. The asymptotical expressions (\ref{cor:14}) coincide with the answers arising in the conventional approach based on the quadratic elasticity theory although they include
the length $\rho_1$ instead of an artificial cut-off radius $\rho_{\rm c}$. The components $\stackrel{(2)}{\si}_{\rho \rho}$, $\stackrel{(2)}{\si}_{\phi \phi}$ behave as $o(\rho^2\log^3\rho)$ at $\rho\to 0$. The stress  $\stackrel{(2)}{\si}_{z z}$ also looks classically at large distances though its long-ranged expression is parametrized by another length $\rho_{\rm m}$ instead of $\rho_{\c}$:
\begin{equation}
\nonumber
  \displaystyle{ \stackrel{(2)}{\si}_{z z}\Bigl|_{\kappa\rho\gg 1}\,\approx\, 2\,\biggl(\nu
\nt\,+\,\Bigl(\frac{b}{2\pi}\Bigr)^2 \mu^3 (1+\nu) C_3
\biggr)\,\biggl(\frac1{\rho^2}\,-\,
\frac2{R^2}\, \log\frac{R}{\rho_\m}\biggr)\,, }
\end{equation}
where $\nu$ is the Poisson ratio (see \cite{def4} for the elastic constant $C_3$). In the gauge approach \cite{def4}, the radii $\rho_1$ and $\rho_{\rm m}$ are expressed through the elastic constants of second and third orders. The appearance of several lengths demonstrates the defect core as a radially layered region without a single sharp boundary.
The continuation of the stresses
is due to the gauge equation (\ref{cor:732}) considered as the non-conventional incompatibility law.

\section{The shear modulus renormalization}
\label{cor:sec3}

The approach of \cite{def4} to description of singularityless dislocations is elaborated in \cite{def5} further for studying the collection of the modified screw dislocations as a thermodynamic ensemble. Specifically, a long enough elastic cylinder pierced by non-singular screw dislocations is studied. The corresponding partition function ${\cZ}$ is given by the functional integral \cite{def5}:
\begin{equation}
{\cZ}\,=\,\int e^{\,\be{\cL}}\,[{\sf {Meas}}]\,,\qquad {\cL}\,\equiv\,{\cL}_{{\rm el}}+{\cL}_{{\rm core}} - i {\cal E}_{{\rm ext}} \,, \label{cor:partf02}
\end{equation}
where $\be$ is inverse of the absolute
temperature, and $[{\sf {Meas}}]$ is
appropriately normalized integration
measure. In the framework of the elasticiy plane problem, ${\cL}$ (\ref{cor:partf02})
includes the standard Lagrangian of linear elasticity
\begin{equation}
{\cL}_{\rm el} =
\frac{-1}{2\mu}\int
({\si}^{\rm b}_{i} + {\si}^{\rm c}_{i})^2\,\d^2x\,,
\label{cor:ellag}
\end{equation}
while the other contributions are:
\begin{equation}
{\cL}_{\rm core}\,=\,\displaystyle{\int
\bigl(\ell\,(\cd_i e_{j}-\cd_j e_{i})^2\,+\,2 e_{i}\,
{\si}^{\rm c}_{i}\bigr)\,\d^2x }\,,\quad
{\cal E}_{\rm ext}\,=\,\displaystyle{\int{\si}^{\rm b}_{i} (\cd_i
u - 2{\cP}_{i})\,{\d}^2x}\,,
 \label{cor:en1}
\end{equation}
where $u\equiv u_3$ and ${e}_{i}\equiv
{e}_{i 3}$ are displacement and total
strain components ($i=1,2$, and $Oz\equiv Ox_3$ is the cylinder axis). Here we use the notations ${\si}^{\#}_{i}\equiv
{\si}^{\#}_{i 3}$ for the stress
components which are independent integration
variables corresponding to, so-called, {\it
background} ($\#$ is $\rm b$) or {\it core}
($\#$ is $\rm c$) contributions.
The non-conventional stress ${\si}^{\rm c}_{i}$ is short-ranged, and therefore it modifies the background one, ${\si}^{\rm b}_{i}$, within the core of the modified dislocation, \cite{def4}. The Lagrangian ${\cL}_{\rm core}$ (\ref{cor:en1}) originates from the Lagrangian (\ref{cor:lagr}). Besides, ${\cP}_{i}$ in
${\cal E}_{\rm ext}$ (\ref{cor:en1}) is the plastic source which is concentrated on cut surfaces bounded by the dislocation lines. It specifies distribution of the background defects.

The functional integral (\ref{cor:partf02}) is estimated in \cite{def5} by means of steepest descent. The stress potential of array of the modified dislocations (\ref{cor:10}) plays the role of the corresponding saddle point
because of the choice of the functional $\cL$
expressed by (\ref{cor:partf02}), (\ref{cor:ellag}),
and (\ref{cor:en1}) (with ${\cP}_{i}$ substituted appropriately).

Therefore the estimate of ${\cZ}$ (\ref{cor:partf02}) leads us to the effective energy ${\CW}=\frac {-1}\be\log\cZ$ of $2\cN$ modified screw dislocations lying along the cylinder axis with unit Burgers vectors ($b=1$) \cite{def5}:
\begin{eqnarray}
{\CW} = \displaystyle{
\frac{- \mu}{4\pi} \sum\limits_{I,
J} \bigl( {\CU}(\kappa |{\bf y}^+_I-{\bf y}^+_J|)\,+\,
{\CU}(\kappa |{\bf y}^-_I-{\bf y}^-_J|)
\,-\,2\,{\CU}(\kappa |{\bf y}^+_I-{\bf y}^-_J|)
\bigr)}\,, \label{cor:eq15}\\
{\CU}(s)\equiv
\log\bigl(\frac\ga 2 s\bigr)\,+\,K_0(s)\,.
\label{cor:eq151}
\end{eqnarray}
The energy ${\CW}$ (\ref{cor:eq15})
demonstrates that the array of the modified screw dislocations is equivalent to the two-dimensional Coulomb gas of unit charges $\pm 1$ characterized by the two-body potential ${\CU}$ (\ref{cor:eq151}) which is logarithmic at large separation but tends to zero for the charges sufficiently close to each other.
Due to condition of ``electro-neutrality'', the number of positive dislocations at the points $\{{\bf
y}^+_I\}_{1\le I\le\cN}$ is equal to the number of negative ones at the points $\{{\bf
y}^-_I\}_{1\le I\le\cN}$.

We consider the grand-canonical ensemble of
the dislocations in the \textit{dipole
phase}, which corresponds to bound pairs of
the dislocations with opposite Burgers
vectors. Define two-point
stress-stress correlation function:
\begin{equation}
\lav{\si}^{\rm }_{i}({\bf x}_1)\,{\si}^{\rm}_{j}({\bf x}_2)\rav\,=\,\displaystyle{{\bf Z}_{\rm dip}^{\1}
\sum\limits_{{\rm n} \& {\rm
p}}\int {\si}_{i}({\bf x}_1)\,{\si}_{j}({\bf
x}_2)\, e^{\,\be{\cL}} [{\sf {Meas}}]} \,, \label{cor:dip1}
\end{equation}
where ${\si}_{i}({\bf x})={\si}^{\rm
b}_{i}({\bf x})+{\si}^{\rm c}_{i}({\bf
x})$ is the total stress distribution, ${\cL}$ is expressed by (\ref{cor:partf02}), (\ref{cor:ellag}), (\ref{cor:en1}), and ${\bf Z}_{\rm dip}$ is the partition function in the dipole approximation.
Besides, $\sum_{{\rm n} \& {\rm
p}}$ is summation over number of dipoles
and over their positions. The
functional integration in (\ref{cor:dip1})
is defined with respect of a given
distribution of the dislocation lines
expressed by ${\cP}_{i}$.

The grand-canonical partition function ${\bf Z}_{\rm dip}$ of the Coulomb system described by the energy (\ref{cor:eq15}) takes the form in the dipole phase \cite{def5}:
\begin{equation}
\begin{array}{rcl}
{\bf Z}_{\rm
dip}&=&\displaystyle{\sum\limits_{{\cN}=0}^\infty\,\,
\frac{1}{{\cN}!} \prod\limits_{I=1}^{\cN} \int\d^2{\bxi}_{I}
\int\d^2{\bta}_{I} }
\,\exp\bigl[-2\be{\cN}\La - \be\CW_{\rm dip}\bigl]\,,
\\[0.4cm]
\CW_{\rm dip} &\equiv& \displaystyle{\sum\limits_{I=1}^{\cN}
{w}({\bta}_{I})\,+\,\sum\limits_{I < J} {w}_{I\!J}}\,,\qquad \beta
w({\eta})\equiv\cK\CU(\kappa \eta)\,,
\end{array}
\label{cor:partf7}
\end{equation}
where $\be=\frac 1T$ is inverse temperature (the Boltzmann constant is unity), ${\cN}$ is the number of dipoles, $\La$ is the chemical potential per dislocation, and $\cK\equiv\frac{\mu\be}{2\pi}$.
Besides, ${w}({\bta}_{I})$ is the energy of $I^{\rm th}$ dipole centered in
${\bxi}_{I}=({\bf y}^+_I+{\bf y}^-_I)/2$ with the dipole momentum ${\bta}_{I}={\bf y}^+_I-{\bf
y}^-_I$. The integration goes over the cylinder's cross-section.
The energy $\CW_{\rm dip}$ in (\ref{cor:partf7}) arises, \cite{def5}, in the dipole approximation from (\ref{cor:eq15}). The energy of interaction between $I^{\rm th}$ and $J^{\rm th}$ dipoles $w_{I\! J}$ in $\CW_{\rm dip}$ (\ref{cor:partf7}) is of the form:
\begin{equation}
\be w_{I\! J} \,\equiv\,- \cK\,
 (\bta_I, {\boldsymbol\cd}_{{\bxi}_{I}}) (\bta_{J}, {\boldsymbol\cd}_{{\bxi}_{J}})\,  \CU(\kappa|{\bxi}_I-{\bxi}_{J}|)\,,
\label{cor:dip312}
\end{equation}
where ${\boldsymbol\cd}_{{\bxi}}$ implies 2-vector $({\cd}_{{\xi}_{1}}, {\cd}_{{\xi}_2}) \equiv
\bigl(\frac{ \cd}{\cd {{\xi}_1}}, \frac{ \cd}{\cd {{\xi}_2}}\bigr)$ and $(\bta_I, {\boldsymbol\cd}_{{\bxi}_{I}})$ is the scalar product of 2-vectors $\bta_I$ and ${\boldsymbol\cd}_{{\bxi}_{I}}$.

The integral in right-hand side of (\ref{cor:dip1}) is calculated in \cite{def5}, and the correlator in the dipole representation of the Coulomb gas takes the form:
\begin{eqnarray}
\lav{\si}_{i}({\bf x}_1) {\si}_{j}({\bf
x}_2)\rav\,=\,\displaystyle{\frac{-\mu}{2\pi \be} \,\cd_{({\bf x}_1)_i}\cd_{({\bf x}_2)_j} \CU(\kappa|{\bf x}_1-{\bf x}_2|)} &&\nonumber
\\[0.3cm]
+\, \displaystyle{{{\bf
Z}^{\1}_{\rm dip}}
\sum\limits_{{\rm n} \& {\rm
p}} {\si}_{i}({\bf x}_1) {\si}_{j}({\bf
x}_2)\,e^{-\be\CW_{\rm
dip}}}\,, && \label{cor:dip111}
\end{eqnarray}
where ${\si}_{i} = \mu \epsilon_{i k}\,\cd_{{x}_k} \phi$, and $\phi\equiv \stackrel{(1)}{\phi}$ is the stress potential (compare with (\ref{cor:10})) of superposition of $\cN$ dipoles. Positions of dipoles are confined within a disk of radius~$R$.

We consider the renormalization of the
shear modulus $\mu$ caused by proliferation
of the dislocation dipoles. The
renormalized $\mu_{\rm ren}$ is defined as
follows:
\begin{equation}
\label{cor:ren1} \displaystyle{\frac{1}{ \mu_{\rm
ren}}\,\equiv\,\frac{\be}{\mu^2 \cS}\,\sum\limits_{i, k =1,
2}\,\iint}\lav{\si}_{i}({\bf x}_1)\,{\si}_{k}({\bf x}_2)\rav\,\d^2{\bx}_1 \d^2{\bx}_2\,,
\end{equation}
where $\cS$ is the cross-section area.
We approximately obtain from (\ref{cor:dip111}) the stress-stress correlation function:
\begin{equation}
\begin{array}{rcl}
&&\lav{\si}_{i} ({\bf x}_1)\,{\si}_{j}({\bf x}_2)\rav\,\approx\,\displaystyle{
\frac{-\mu}{2\pi \be}}\,\Bigl(\cd_{({\bf x}_1)_i}\cd_{({\bf x}_2)_j} \CU(\kappa |{\Dl}{\bx}|)\Bigr. \\[0.4cm]
&&- \Bigl. \displaystyle{\sum \limits_{{\tilde n}=1}^\infty
(-\be\mu d)^{{\tilde n}}
\bigl(\epsilon_{i k}
\epsilon_{j l}\,\cd_{({\bf x}_1)_k}
\cd_{({\bf x}_2)_l}\bigr)
\Bigl[\CU(\kappa |\Dl{\bf
x}|)\,+\frac{\kappa |\Dl{\bf x}|}{2}\,K_1(\kappa
|\Dl{\bf x}|)\Bigr]  } \Bigr)\,. \label{cor:eq27}
\end{array}
\end{equation}
The term given by ${\tilde n}=1$ in right-hand side of (\ref{cor:eq27}) coincides with that obtained in \cite{def5} in the case of non-interacting dipoles.
The contributions at ${\tilde n}\ge 2$ in right-hand side of (\ref{cor:eq27}) are due to the dipole-dipole coupling (\ref{cor:dip312}). One uses (\ref{cor:ren1}), (\ref{cor:eq27}) and obtains:
\begin{equation}
\displaystyle{\frac{1}{\mu_{\rm ren}}\,=\,\frac{1}{\mu}\,
{\cC}_1(\kappa R)\,+\,\frac{\be d}{1+ \mu\be d}\,{\cC}_2(\kappa R)}\,,
\label{cor:ren2}
\end{equation}
where $d$ is proportional to mean area
covered by the dipoles (more precisely, $\be \mu d \equiv \pi\cK \l {\bta}^2 \r\bar N$, where $\l {\bta}^2 \r$ is mean square of the dipole momentum, and $\bar N$ is average dipole density, \cite{def5}). The functions ${\cC}_1(\kappa R)$ and ${\cC}_2(\kappa R)$ are given by the modified Bessel functions:
\begin{equation}
\begin{array}{rcl}
\displaystyle{{\cC}_1(\kappa R)}&
=&\displaystyle{1\,-\,2 K_1(\kappa
R) I_1(\kappa R)}\,,
\\[0.4cm]
\displaystyle{{\cC}_2(\kappa R)}
&=& {2\,-\,2 I_1(\kappa R)\bigl(K_1(\kappa
R)\,-\,\kappa R\,K_1^{\prime}(\kappa R)\bigr) }\,,
\nonumber
\end{array}
\end{equation}
and $K_1^{\prime}(z)=\frac{\d}{\d z}K_1(z)$. The influence of the cores causes the size-dependence of the functions ${\cC}_1(\kappa R)$ and ${\cC}_2(\kappa R)$.
Equation (\ref{cor:ren2}) demonstrates that the shear modulus $\mu_{\rm ren}$ depends on the dimensionless
ratio $R/\kappa^{\1}=\kappa R$ of two
lengths characterizing the sample's
cross-section and the dislocation core
sizes. The coefficients ${\cC}_1(\kappa R)$ and ${\cC}_2(\kappa R)$ both are
positive and less than unity though tend to unity at $\kappa R\to \infty$:
\begin{equation}
\displaystyle{{\cC}_1(\kappa R)\,
\approx\,
1\,-\,\frac{1}{\kappa R}\,+\,\dots\,,\qquad
{\cC}_2(\kappa R)\,
\approx\,
1\,-\,\frac{3}{2 \kappa R}\,+\,\dots}\,,
\nonumber
\end{equation}
where the ellipsis imply the terms ${\cal O}((\kappa R)^{-2})$.

The dependence on the size-parameter $\kappa R$ in Eqs.~(\ref{cor:ren2}) displays the effect of the non-conventional dislocation solution on the shear
modulus near the melting transition. As it has been experimentally confirmed in \cite{grun}, properly normalized Young modulus tends to the universal value $16\pi$ at $T\to T_c^-$ (see \cite{kl12}). The present approach demonstrates that the renormalized shear modulus deviates from a multiple of $\pi$ due to the smoothed out character of the dislocations:  \begin{equation}
\frac{\mu_{\rm
{ren}}(T^-_c)}{{T_c}}\approx \frac {8\pi}{{\cC}_1(\kappa R)}
\begin{CD}@>> \kappa R\gg 1>\end{CD}\,8\pi \,, \qquad d\ll 1\,.
\label{cor:ren100}
\end{equation}

\section{Discussion}

The results obtained should be applicable
to nanotubes and nanowires with
comparable $R$ and $\kappa^{-1}$.
With regard at the results of \cite{grun, capac} concerning the colloidal crystals, it is hopeful
that experimental nanophysics could
provide us opportunities of verification of the relations
of the type of Eqs.~(\ref{cor:ren2}) and (\ref{cor:ren100}) near the melting point.\footnote{Defect mediated melting attracts attention in various systems, \cite{baym, superc, dipol}.} The colloidal crystals have also been mentioned in \cite{rab} among other candidates for observing the effects due to the elastic constants renormalization.
Further studies of the effects
due to the singularityless character of the dislocations look attractive as far as nano-materials and nano-physics are concerned.

\section*{Acknowledgments}

Partially supported by the Russian Science Foundation (grant no. 14-11-00598).


\begin{thebibliography}{99}

\bibitem{kl11}
         H. Kleinert, {\it Gauge Fields in Condensed Matter}.
         Vol. I (World Scientific,
         Singapore, 1989).

\bibitem{kl12}
         H. Kleinert, {\it Gauge Fields in Condensed Matter}.
         Vol. II (World Scientific,
         Singapore, 1989).

\bibitem{mult}
         H.~Kleinert, \textit{Multivalued Fields in Condensed Matter, Electromagnetism, and Gravitation } (World Scientific,
         Singapore, 2008).

\bibitem{kat}
         M.~O.~Katanaev, I.~V.~Volovich, {\it Ann. Phys.} {\bf 216}, 1 (1992).

\bibitem{mok}
         M. O. Katanaev, {\it Physics--Uspekhi} {\bf 48}, 675 (2005).

\bibitem{mok1}
         G. de Berredo-Peixoto, M. O. Katanaev, {\it J. Math.
Phys.} {\bf 50}, 042501 (2009).

\bibitem{katmeth}
         M.~O.~Katanaev, \textit{Geometrical Methods in Mathematical Physics}, {\sl  arXiv:1311.0733v3} {\sl [math-ph]}.

\bibitem{nelson}
     D.~R.~Nelson,
     \textit{Defects and Geometry in Condensed Matter Physics}
     (CUP, Cambridge, 2002).

\bibitem{hugh}
      Gil Young Cho, O. Parrikar, Yizhi You, R. G. Leigh, T. L. Hughes,
      \textit{Phys. Rev. B} \textbf{91}, 035122 (2015).

\bibitem{liu}
        Ke Liu, J.~Nissinen, Z.~Nussinov, R.-J.~Slager, Kai Wu, J.~Zaanen,
         \textit{Phys. Rev. B}
         \textbf{91}, 075103 (2015).

\bibitem{vol1}
  G.~E.~Volovik, M.~A.~Zubkov, \textit{Ann. Phys. (NY)} {\bf 356}, 255 (2015).

\bibitem{val}
         M. C. Valsakumar, D. Sahoo, {\it
         Bull. Mater. Sci.} \textbf{10}, 3 (1988).

\bibitem{ed1}
         D. G. B. Edelen, {\it Int. J. Engng Sci.} \textbf{34}, 81 (1996).

\bibitem{laz2}
         M. Lazar, \textit{J. Phys. A:
         Math. Gen.} \textbf{35}, 1983 (2002).

\bibitem{def3}
     C. Malyshev, {\it Ann. Phys. (NY)} \textbf{286}, 249 (2000).

\bibitem{def4}
     C. Malyshev, {\it J. Phys. A: Math. Theor.} \textbf{40}, 10657 (2007).

\bibitem{str1}
 M.~Caselle, P.~Grinza, N.~Magnoli,
  {\it J. Stat. Mech.} \textbf{0611}, P11003 (2006).

\bibitem{str2}
E.~R.~Bezerra de Mello, V.~B.~Bezerra, A.~A.~Saharian, H.~H.~Harutyunyan, {\it Phys. Rev. D} \textbf{91}, 064034 (2015).

\bibitem{vol2}
     G.~E.~Volovik, \textit{Phys. Scr. T} {\bf 164}, 014014 (2015).

\bibitem{def5}
     C. Malyshev, {\it J. Phys. A: Math. Theor.} \textbf{44}, 285003 (2011).

\bibitem{aopd}
     C. Malyshev, {\it Ann. Phys. (NY)} \textbf{351}, 22 (2014).

\bibitem{holz}
         A. Holz, J.~T.~N.~Medeiros,
         \textit{Phys. Rev. B} \textbf{17}, 1161
         (1978).

\bibitem{nel1}
         D.~R.~Nelson, \textit{Phys. Rev. B} \textbf{18}, 2318 (1978).

\bibitem{nel12}
         D.~R.~Nelson, B.~I.~Halperin,
         \textit{Phys. Rev. B} \textbf{19}, 2457
         (1979).

\bibitem{nel13}
         A. P. Young, \textit{Phys. Rev. B} \textbf{19}, 1855 (1979).

\bibitem{grun}
         H. H. von Gr\"unberg, P. Keim,
         K. Zahn, G. Maret, \textit{Phys.
         Rev. Lett.} \textbf{93}, 255703 (2004).

\bibitem{rab}
         S. Panyukov, Y. Rabin, \textit{Phys. Rev. B} \textbf{59}, 13657 (1999-I).

\bibitem{ga}
         M. Yu. Gutkin, E. C. Aifantis,
         \textit{Scripta Mater.} \textbf{40}, 559
         (1999).

\bibitem{capac}
   S.~Deutschl\"ander, A.~M.~Puertas, G.~Maret, P.~Keim, \textit{Phys. Rev. Lett.} \textbf{113}, 127801
   (2014).

\bibitem{baym}
     S.~A.~Gifford, G.~Baym, \textit{Phys. Rev. A} \textbf{78}, 043607 (2008).

\bibitem{superc}
     T.~Saiki, R.~Ikeda, \textit{Phys. Rev. B} \textbf{83}, 174501 (2011).

\bibitem{dipol}
    H. Kleinert, {\textit{Europhys. Lett.}} \textbf{102}, 56002 (2013).

\end{thebibliography}
\end{document}